\documentclass{article}
\usepackage{amsmath,amssymb,mathrsfs,amsfonts,dsfont}
\usepackage[utf8]{inputenc}
\usepackage{graphicx}
\usepackage[numbers]{natbib}
\usepackage{tikz}
\usetikzlibrary{positioning}
\usepackage[T3,OT1]{fontenc}
\DeclareSymbolFont{tipa}{T3}{cmr}{m}{n}
\DeclareMathAccent{\invbreve}{\mathalpha}{tipa}{16}

\newlength{\defbaselineskip}
\newcommand{\setlinespacing}[1]%
           {\setlength{\baselineskip}{#1 \defbaselineskip}}

\newcommand{\singlespacing}{\setlength{\baselineskip}{\defbaselineskip}}

\begin{document}
\begin{center} {\LARGE\textbf{How device-independent approaches change the meaning of physical theory}}
\vskip 2em
{\large \bf Alexei Grinbaum} \\
{\it CEA-Saclay/IRFU/LARSIM, 91191 Gif-sur-Yvette, France
\par Email alexei.grinbaum@cea.fr}
\vskip 1em 
\end{center}


\abstract{\noindent Dirac sought an interpretation of mathematical formalism in terms of physical entities and Einstein insisted that physics should describe ``the real states of the real systems''. While Bell inequalities put into question the reality of states, modern device-independent approaches do away with the idea of entities: physical theory may contain no physical systems. Focusing on the correlations between operationally defined inputs and outputs, device-independent methods promote a view more distant from the conventional one than Einstein's `principle theories' were from `constructive theories'. On the examples of indefinite causal orders and almost quantum correlations, we ask a puzzling question: if physical theory is not about systems, then what is it about? The answer given by the device-independent models is that physics is about languages. In moving away from the information-theoretic reconstructions of quantum theory, this answer marks a new conceptual development in the foundations of physics.}

\section{Introduction}

Often hailed as a ``second quantum revolution''~\cite{AspectBell}, the introduction of correlation inequalities by John Bell~\cite{Bell1} inaugurated a conceptual development whose significance took several decades to be fully appreciated. We submit that this revolution reaches a surprising summit with the development of device-independent approaches and model-independent physics, supporting a dramatically new view of physical theory.

Quantum mechanics describes the evolution of a system under a particular Hamiltonian and the results of measurements operated on this system by the observer. The concept of observer is external to the theory. Whatever its physical constitution, the observer's only role is to choose a measurement setting and register the result of the observation: an operational approach. Correlations between the observer's choices and results are intuitively taken to be mediated by information carriers: physical systems. On one view, systems are ``lines'' or ``wires'' between ``boxes'' in symbolic diagrams connecting various operations on the observer's information---a conception that leads to ``new modes of explaining physical phenomena''~\cite{coecke_pict,coecke_NJ,coecke_struct}.
The old explanatory mode, on the contrary, takes systems to be constituted through separation from non-systems (measurement devices or the environment): a system is a bouquet of relevant degrees of freedom jointly denoted by a single name. That such a division, although it is not a \textit{definition} (more on this in Section~\ref{sect_principleconstructive}), enables \textit{explanation} is an idea with a long philosophical history ($\delta\iota\epsilon\tilde{\iota}\lambda\epsilon\nu$ from $\delta\iota\alpha\iota\rho\acute{\epsilon}\omega$, to take apart, Plato \textit{Timaeus} 41d). We argue, firstly, that the old explanatory mode does not apply to device-independent approaches. Secondly, in the new explanatory mode systems become auxiliary concepts and, like any accessory tool, have limited utility. Still occasionally employed in the literature, they represent little more than a counterintuitive and unhelpful remnant of the old regime. Our examples will show that, while it is not always outright wrong, thinking about physics in terms of systems sometimes hinders rather than facilitates understanding. It is significant, then, that the new explanatory mode can produce a physical theory that does not refer to systems at all.

In quantum mechanics, it is assumed that a measurement setting is chosen in earnest, i.e., the observer trusts the system to be constituted of precisely the degrees of freedom described by the theory. What the system is, is known in advance and is correct. For example, if one performs a binary measurement of photon polarization, then one expects \textit{a priori} that the measurement device will indeed measure photons. This trust in preparation devices is usually not subject to theoretical scrutiny, yet it is in principle---and often experimentally---unfounded.

The problem of trust contains a further aspect. If the distinction between a system and a measurement device is fixed within one laboratory, then it is usually taken for granted that all other laboratories, should they come to observe the processes in the first one, will make the same distinction along the same separation line. The identity of the system does not depend on the observer; only its state may vary in relation to the observer's choice of measurement. The ``Wigner's friend'' gedankenexperiment~\cite{WignerMind} assumes that different observers will agree on system identification but disagree on state ascriptions. It is understandable that this agreement may be a matter of unassailable trust between friends; it has been put into question and studied mathematically only recently~\cite{grinbKolm,pienaar}.

Absence of trust is a concern that quantum cryptography is designed to address. It has tools for working with systems of ``unspecified character''~\cite{PhysRevLett.106.250404} or ``unknown nature''~\cite{PhysRevA.80.062327}. A device-independent approach employs such tools: it is
a theoretical investigation performed without relying on the knowledge of the laws governing the systems' behaviour. A conventional `device' refers here to any process or apparatus described by an operational theory, whether classical or quantum, which is explicitly designated. In this sense, not only a conventional optical table but something as strange as a closed timelike curve~\cite{DeutschCTC,BennettCTC} or a Malament-Hogarth spacetime~\cite{EarmanNorton,Hogarth} may be seen as a device. This terminology was first introduced by Mayers and Yao~\cite{MayersYao}, who developed device-independent quantum cryptography with imperfect sources. Their suggestion was to render, through a series of tests, an untrusted but ``self-checking'' source equivalent to an ideal one that can be trusted \textit{a priori}. These tests do not rely on the degrees of freedom pertinent to the system or, to put it differently, on our knowledge of the physical theory that describes their evolution. They only involve inputs and outputs at two separate locations: a device-independent protocol (Section~\ref{sect_def}). Over the years quantum cryptography has developed an array of such methods for dealing with adversaries which, via action upon sources, effectively turn systems into untrusted entities. Device-independent protocols are important for randomness generation~\cite{ColbeckThesis,PironioNature}, quantum key distribution~\cite{barrett_no_2005}, estimation of the states of unknown systems~\cite{PhysRevA.80.062327}, certification of multipartite entanglement~\cite{PhysRevLett.106.250404}, and distrustful cryptography~\cite{MassarPironio}.

Some of these cryptographic protocols have found a broader use in quantum information, e.g. device-independent tests are performed on Bell inequalities, on the assumption that superluminal signaling is impossible~\cite{Bancalbook}, or on the existence of a predefined causal structure (Section~\ref{sect_causal}). But the import of device-independent methods extends even further. Device-independent methods convert the usually implicit trust of the observer into a theoretical problem. By doing so, they erase one of the main dogmas of quantum theory: that it deals with systems. To appreciate the significance of this shift, we compare it with another paradigmatic change captured by Einstein in the form of a distinction between principle and constructive theories (Section~\ref{sect_principleconstructive}).

This shift is not only due to the import of device-independent methods from quantum cryptography into general quantum physics. If these methods have indeed triggered the development, the latter had been prepared by the reconstructions of quantum theory (Section~\ref{sect_reconstruct}). Operational axiomatic approaches to quantum mechanics focus on the inputs and outputs of the observer: a ``box'' picture. The postulates that successfully constrain the box to behave according to the rules of quantum theory become our best candidates for fundamental principles of Nature. In a device-in\-depen\-dent approach, such postulates are also at work: they are the only content of physical theory along with the inputs and the outputs of the parties.

\begin{figure}\centering
\includegraphics[scale=0.4]{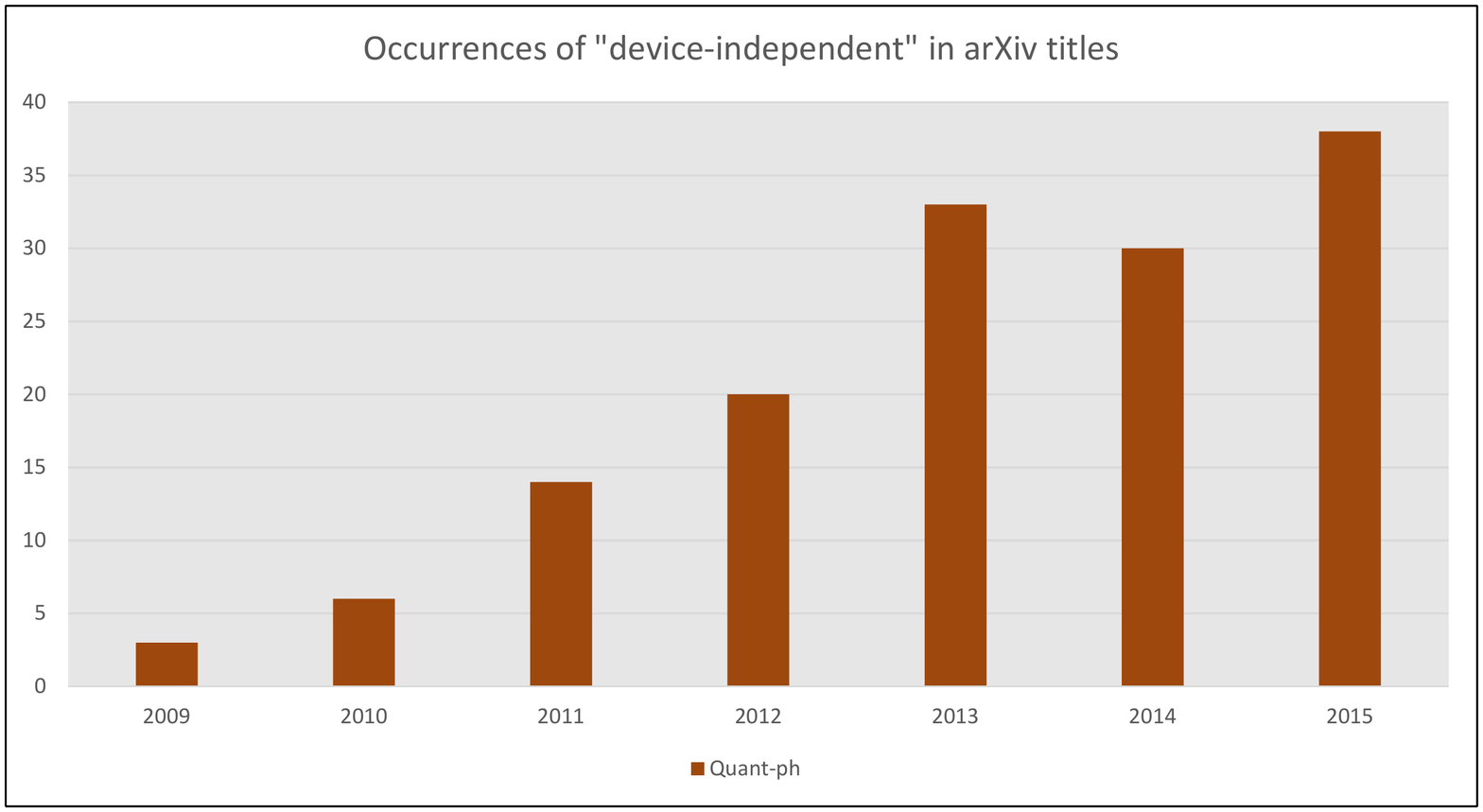}\label{fig_stats}
\caption{Occurrences of the term ``device-inde\-pen\-dent'' in the titles of arXiv physics preprints.}
\end{figure}

Incompatible with the old explanatory mode, device-in\-depen\-dent models typically do not meet the conditions for the emergence of robust theoretical constituents corresponding to real objects. By allowing no room for systems, they inaugurate the obsolescence of this elementary building block: a theory may contain no systems but remain physical. The spread of this view from quantum cryptography to general quantum physics (Figure~\ref{fig_stats}) raises a question of meaning: if physical theory is not about systems, what is it about? This requires a philosophical (Section~\ref{sect_onto}) as well as a mathematical (Section~\ref{sect_almost}) investigation. Device-independent models suggest a possible answer: physical theory is about languages (Section~\ref{sect_lang}). Not only is such a theory possible; the spread of device-independence shows that it may become routine. Perhaps it indicates the right direction for moving beyond quantum theory.

\section{Physics in a box}\label{sect_def}

\begin{figure}
\centering
\includegraphics[scale=0.5]{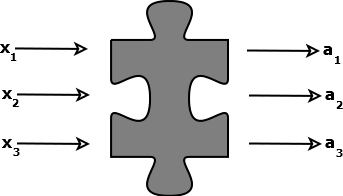}
\caption{In the case of $n=3$ parties, physics is fully contained in the probabilities $\mathbf{p}=P(a_1 a_2 a_3|x_1 x_2 x_3)$.}\label{boxes}
\end{figure}

Device-independent models are defined as a set of $n$ parties, each of which `selects' a measurement setting or `places' an input value $x_1\in \mathcal{X}_1,\ldots ,x_n\in \mathcal{X}_n$ respectively, and `subsequently' `obtains' an output value or a measurement result $a_1\in \mathcal{A}_1,\ldots , a_n\in \mathcal{A}_n$. The sets $\mathcal{X}_1,\ldots,\mathcal{X}_n$ and $\mathcal{A}_1,\ldots,\mathcal{A}_n$ are alphabets of finite cardinality. The verbs used in these expressions merely convey an operational meaning of the inputs and outputs; they do not imply that any party exercises free will or has conscious decision-making procedures. The term `subsequently' introduces a local time arrow pointing from each party's input to its output. Although such local time arrows seem quite intuitive, in full generality they need not be assumed either. A fully general setting requires, therefore, that absolutely nothing be postulated about the way inputs are transformed into outputs, except two conditions: a) these two types of data are clearly distinguished; b) the process of transformation is physical. Physics is contained in the probability distribution $\mathbf{p}=P(a_1,\ldots,a_n|x_1,\ldots,x_n)$ (Figure~\ref{boxes}).

All device-independent models studied in the literature introduce further constraints on $\mathbf{p}$. The most frequent one is the no-signalling principle: a choice of measurement by one party must not influence the statistics of the outcomes registered by a different party. Mathematically, the distribution $\mathbf{p}$ is non-signalling if and only if all one-party marginal probabilities are functions of their respective inputs $x_i$: \begin{equation}P(a_i|x_1,\ldots,x_n)=P(a_i|x_i).\end{equation} Although very common, this assumption is not universal: when device-in\-depen\-dent methods are used to test general causal inequalities, the impossibility of signalling is not a prerequisite~\cite{Baum1}.

It is possible to argue that the property of device-independence was already apparent in Bell's own formulation of his inequalities~\cite{Bell1}. However, the first proper \textit{model} featuring non-signalling and device-independence is to be found in the work of Popescu and Rohrlich~\cite{popescu}. A non-local, or Popescu-Rohrlich (PR), box describes unknown processes which connect the inputs $x,y\in\{0,1\}$ and the outputs $a,b\in\{0,1\}$ of two parties according to the joint distribution
\begin{equation}\nonumber
P(ab|xy) = \left\{\begin{array}{l@{\quad}l} 1/2:&
a+b={xy} \mod 2\\
0:& \mathrm{otherwise.}
\end{array} \right.
\end{equation}
The no-signalling constraint implies that, while a PR-box is designed to go beyond quantum theory, it nevertheless respects the laws of special relativity. Its device-independent non-local structure accommodates a violation of the Tsirelson bound~\cite{Tsirel} by reaching the maximum amount of correlations in the CHSH inequality. Since PR-boxes allow for more-than-quantum (often called \textit{postquantum}) correlations, they cannot be built experimentally given the current state of knowledge. There exist, however, experimental approximations with the no-signalling condition weakened through a coordinated choice of measurement settings~\cite{PhysRevLett.113.100401,Bourenanne} or postselection~\cite{PhysRevA.75.022102}.

Hailed as a ``very important recent development''~\cite{popescu2014}, device-in\-depen\-dent models are characterized by the absence of assumptions about the internal workings of the box. Its `interior' is not described by a particular physical theory. The box is unknown territory which, since it is  assumed to be of interest for physical theory, is also a territory of science. The entire setup belongs within the boundaries of physics (the workings of the box are not miracles) and, at the same time, it opens a possibility to redefine these very boundaries. It may be the case that $\mathbf{p}$ is consistent with the predictions of an available physical theory, but if this is not so, then the meaning of physical theory is appropriately widened to include the correlations realized by the box.

\section{Example: Causal orders}\label{sect_causal}

No-signalling is the most commonly used condition on $\mathbf{p}$. Other, usually more concrete examples are also formulated as informa\-tion-\-theor\-etic con\-straints, e.g., a condition on the security of bit commitment~\cite{MassarPironio}. While they take the box closer to quantum theory, such assumptions still leave enough room for models beyond quantum mechanics, giving quantum theory a place in a broader landscape. Research on `indefinite causal orders' does not rely on a constraint on $\mathbf{p}$ imported from quantum communication. It explores another surprising feature of device-independence: the absence of global temporal order between the inputs and the outputs associated with different parties. Each party, for sure, can draw an arrow pointing from its input to its output, the latter always succeeding the former in this party's local frame of reference. While such local time axes are well-defined, Chiribella~\cite{Chiri_causal} following Hardy~\cite{hardy_probability_2005,hardy_towards_2007} suggested that there may not exist a global notion of time. His work pursued a device-dependent approach, whereby local transformations were taken to be quantum but the big Hilbert space of all parties contained no information on causal relations among them. A mathematical formalism called `process matrix' was introduced by Oreshkov, Costa and Brukner to deal with such situations~\cite{oreshkov_quantum_2012}, leading to a set of further studies~\cite{Brukner2014,Brukner_rev,Brukner2015_1,Brukner2015_2,Issam1,Amin1,Feix15}.

\begin{figure}
		\centering
		\begin{tikzpicture}
			\node[draw,rectangle,minimum width=1cm,minimum height=1cm] (S) {$\mathcal{E}$};
			\draw[->] (S.150) -- ++(-0.2,0) node [left] {$A$};
			\draw[<-] (S.210) -- ++(-0.2,0) node [left] {$X$};
			\draw[->] (S.90) -- ++(0,0.5) node [above] {$O_S$};
			\draw[<-] (S.270) -- ++(0,-0.5) node [below] {$I_S$};
		\end{tikzpicture}\label{amin_picture}
\caption{A party is fully defined by the input variable $X$ and the output variable $A$ linked by the map $\mathcal{E}$. The theory does not require any mention of a physical system $S$ that is first received from the environment and then returned to it. Adopted with modifications from~\cite{baumeler_2}.}
\end{figure}
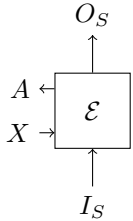

In a device-independent approach, a party is defined, not by a local Hilbert space, but by two random variables: an input $X$ and an output $A$, and a map $\mathcal{E}$ between them. It is conventional among authors to mention ``a physical system $S$ that a party receives from the environment and a physical system that is returned to the environment''~\cite{baumeler_2} and to adduce it to the transformation $\mathcal{E}$ (Figure~\ref{amin_picture}). This is not due to any theoretical necessity: the language merely appends an interpretation of the mathematical formalism. Three strange consequences follow. Taken together, they indicate that the notion of system is not a fundamental ingredient of the device-independent model of indefinite causal orders.

First, the assumption about a physical system first entering, then leaving the laboratory can be rephrased as a condition that each party interacts with the `environment' or the `physical medium' only once:
\begin{quote}
[Alice and Bob] both open their lab, let some physical system in, interact with it and send a physical system out, only once during each run of the experiment.~\cite{Branciard15}
\end{quote}
The notion of environment here involved is a peculiar one. Described as a whole by the process matrix, it lies outside space-time. `Environment' is used as a name for a holistic atemporal medium said to supply a system to a party. According to the old explanatory mode of physics, a system is to be constituted through separation from the environment; however, no separation is possible from this one. If the relevant degrees of freedom could be divided from the irrelevant ones in this `environment', then the degrees of freedom pertaining to the system would remain identifiable as such throughout the experiment. The idea of maintaining ($\sigma\upsilon\nu\acute{\epsilon}\chi o\nu\tau o\varsigma$) this separation, or holding together ($\sigma\upsilon\nu\acute{\alpha}\gamma o\nu\tau o\varsigma$) the degrees of freedom that constitute the system, is key to providing cohesion ($\sigma\upsilon\gamma\kappa\rho\alpha\tau o\invbreve{\upsilon}\nu\tau o\varsigma$) of whatever is separated or divided from something else (Numenius 4b). This is a temporal idea, at least for the time of the experiment: if a system melts down, or is broken up, or gets absorbed inside the laboratory, then it ceases to be. The `holding together' of the degrees of freedom happens in the laboratory's time, but this local time arrow does not extend to the holistic environment described by the process matrix. The latter is not in space-time, hence the impossibility to use a global medium to maintain separation outside the lab, or to maintain anything \textit{tout court}. As a result, the framework of indefinite causal orders cannot accommodate a notion of system which `enters' or `leaves' the lab from the environment.

Second, in some circumstances, `systems' in the process matrix framework may `enter' the same local laboratory twice: a situation that never occurs to a physical object. Third, both quantum bipartite and classical multipartite processes without predefined causal order are logically consistent and therefore allowed by the theory~\cite{baumeler16}. Causal correlations between the inputs and the outputs form a polytope that lies in a larger, logically consistent set that includes non-causal correlations. The latter can be shown to violate a causal inequality: an analog of Bell inequality that permits to distinguish between mixtures of predefined causal orders and its genuine absence. In the quantum framework, violations arise already with two parties~\cite{oreshkov_quantum_2012}. With three or more parties, it becomes possible to reach a non-causal point in the larger polytope using only classical probability theory~\cite{Amin1}. This is a surprising finding, implying that even a classical device-independent framework cannot always be interpreted as a description of physical systems entering and leaving the laboratories.

These three reasons underscore the difficulty to employ the notion of system in a device-independent approach. Another line of research gives weight to this conclusion by addressing the following point: if a system `traverses' the laboratory `from' the input `towards' the output, one should be able to tell its history, i.e., provide a list of events that occurred to the system in the laboratory. It is, however, impossible; the most radical manifestation of which is the ``quantum liar'' paradox~\cite{QLiar}. Based on postselection, this example showcases a paradoxical conclusion that a future measurement may take an active part in the formulation of the system's past history. Thus, a nearly automatic phrase: ``a system is prepared at the start of an experiment,'' if taken seriously, produces an array of counterintuitive consequences. This phrase may best be abandoned.

Three versions of the same condition pertaining to experimental investigation each characterize a particular aspect of physics.
\begin{description}
\item[a)] To underline the instrumental or the operational aspect, one chooses as a primitive a unique run of the experiment fully described by an input and an output.
\item[b)] To put an emphasis on theory as opposed to experiment, the parties (or the laboratories) are defined, not by spatial arrangements of instruments, but by an input and an output.
\item[c)] On the interpretational side, it is commonly assumed that a physical system enters a laboratory and then leaves it.
\end{description}
While they seem complementary, these three different readings may not be equally necessary. The instrumentalism of a) takes a fast route to establishing the mathematical formalism of physical theory. It is then sufficient to follow b) for a purely formal investigation as it contains all the information needed for doing calculations. A `system' in c) is a mere interpretative device that runs into difficulties when brought to light. It is the least necessary assumption and, unable to give a hand conceptually to a) and b), it often becomes counterproductive as it mires the meaning of physical theory. A theory can be provided by a) and b), without c), containing no systems while staying physical.

\section{Relation to `principle theories'}

\label{sect_principleconstructive}

The question about what amounts to a physical theory is usually debated in the light of a well-known distinction, drawn by Einstein in 1919~\cite{ein19}, between constructive and principle theories. A paradigmatic example of principle theories is Einstein's own special theory of relativity: its entire edifice is derived from simple postulates that reflect abstract universal principles of Nature, not the laws of behaviour of a particular kind of matter. As Einstein noted, such principles serve to ``narrow the possibilities''~\cite{ein_galison}. The same `narrowing of possibilities' is achieved by introducing constraints on $\mathbf{p}$ in the device-independent approach. This suggests a possible link between the meaning of the latter and Einstein's distinction.

Special relativity is not a constructive theory, i.e., it remains mute on the issue of material constitution of the rods and clocks that act as its measurement devices. Einstein believed that this lack of constructivity was a disadvantage and, consequently, principle theories did not offer a satisfactory understanding of physics ~\cite{bt}. He kept hoping that a constructive theory could provide a better understanding of Nature: ``When we say we have succeeded in understanding a group of natural processes, we invariably mean that a constructive theory has been found which covers the processes in question``~\cite{ein19}. But Einstein's desire to obtain a constructive theory as a replacement of his principle-based special relativity never came to be realized. It is tempting to speculate that device-dependent physics describing concrete physical systems will follow the same destiny as constructive theories. If, despite Einstein's wish, no constructive theory has materialized as a replacement of special relativity, it is not impossible to imagine that our intuitive desire to `fill the box' with physical systems for the purposes of better explaining physics is as illusory. The device-independent approach might stay as a legitimate way of doing physics, without any need to `fill the box,' much in the same sense as principle-based special relativity has not been surpassed by any constructive theory.

Device-independent approaches inaugurate a bigger shift from concrete physics than Einstein's principle theories. The latter assume, just as constructive theories do too, that the elementary building blocks of physical theory are physical systems. Constructive theories put a direct emphasis on this assumption as they begin their development from certain elementary material constituents. Theoretical entities are, in this case, mere formal representations of real objects. Principle theories achieve a similar conclusion from the opposite direction, by postulating general principles in order to derive a theory of entities constrained by them. Physical systems are now theoretical constructs to be put in correspondence with the real objects. None of the two types of theories includes a possibility that physical theory may not contain any entities, whether real or theoretical, and may not seek to develop a notion of system. Einstein certainly did not envision such a physical theory, be it principle or constructive. In philosophical remarks written in 1949, he states that, even if the directly perceived and consequently conceptualized sensory impressions are deceitful candidates for being a part of reality, theoretical entities are nevertheless necessary in a physical theory:
\begin{quote}
  A basic conceptual distinction, which is a necessary prerequisite of scientific and pre-scientific thinking, is the distinction between ``sense-impressions'' on the one hand and mere ideas on the other. There is no such thing as a conceptual definition of this distinction (aside from circular definitions). Nor can it be maintained that at the base of this distinction there is a type of evidence, such as underlies, for example, the distinction between red and blue. Yet, one needs this distinction in order to be able to overcome solipsism. Solution: we shall make use of this distinction unconcerned with the reproach that, in doing so, we are guilty of the metaphysical ``original sin.''~\cite{EinRem49}
\end{quote}
As they reach a new level of abstraction from concrete material reality, device-independent approaches surpass Einstein's view that the ``sense-impressions,'' now in the form of the inputs and the outputs, must be connected with the idea of system. But Einstein's remarks still carry a deep insight, \textit{viz.}, that a definition of system cannot be given rigorously. Fundamental theoretical notions cannot be defined without circularity only based on phenomenal data, because our reading of data is itself theory-laden; this is why a system does not have a \textit{definition}. At the same time, the presence of theoretical entities enables physical \textit{explanation}. According to Einstein, it is necessary for \textit{any} physically explanatory account---a claim to be challenged by device-independent approaches. In a markedly Kantian development, Einstein uses the argument from usefulness to support his position:
\begin{quote}
It is also the presupposition of every kind of physical thinking. Here too, the only justification lies in its usefulness. We are here concerned with ``categories'' or schemes of thought, the selection of which is, in principle, entirely open to us and whose qualification can only be judged by the degree to which its use contributes to making the totality of the contents of consciousness ``intelligible.''~\cite{EinRem49}
\end{quote}
The history of Kantian interpretations of physical theories~\cite{desp,Bitbol,Petitot} teaches us that, on many an occasion, what had been taken \textit{a priori} by earlier thinkers, later on became a potentially variable object of scientific investigation: a \textit{relativized a priori}~\cite{Friedman}. Device-independent approaches accomplish precisely such a relativization of theoretical entities previously assumed by Einstein to be necessary for intelligibility. They exhibit a case in which, contrary to Einstein's presupposition, the notion of system can become unhelpful and it is possible to understand a physical model without recourse to this idea.

\section{Relation to the reconstructions of quantum theory}\label{sect_reconstruct}

The introduction of principle theories by Einstein and a vision of mathematical physics promoted by the Hilbert program have both contributed to the rise of quantum axiomatics. This line of research began with the proposal of quantum logic by von Neumann and Birkhoff in 1935~\cite{bvn}, showcasing a change in the foundational attitude from a physical enquiry dealing with real objects to a mathematical formalism that only contains theoretical entities. In a departure from the Hilbert space quantum mechanics, von Neumann ``made a confession'' in a letter to Birkhoff that he did not believe in the Hilbert spaces any more~\cite{vNletters}. To describe physical systems in a different way, a correspondence was to be established between measurements and a projective-geometric structure isomorphic to an orthomodular lattice. Several decades of research along these lines in quantum logic yielded multiple proposals for the axioms of quantum theory. Orthodox quantum logic was followed by a reconstruction program focused on the operational meaning of quantum theory~\cite{grinbjps}. In contrast to the previous, heavily mathematical axioms, reconstructions sought to identify a small set of principles with a clear physical meaning. With no exceptions, these axiomatizations contained a postulate about the subsystems and the composition rule (e.g., in~\cite{hardy_quantum_2001}), whose function was to put a limit on the amount of correlations that can be reached by the subsystems. Postulates of this kind reduce the maximally allowed set of bipartite or multipartite correlations down to the quantum bound; this can only be achieved, however, if what needs to be derived is already known. The reconstruction program of quantum theory, therefore, sought to reconstruct an already existing theory.

At the same time, composition rules take extra meaning in a more general device-independent approach that goes beyond reconstruction. Imagine that no subsystems are introduced but only a limit on the correlations. This device-independent setup operates with the inputs and the outputs of the parties while it contains no notion of system. Now, the available limit on correlations, acting as a constraint on $\mathbf{p}$, is used to derive the conditions under which a notion of system would become meaningful. Systems emerge, then, as a result of some principles, which are usually formulated in the information-theoretic language.
Device-independent approaches drive home the importance of such principles: quantum theory appears as one among several possible information-theoretic models. Its meaning in this context has a fainter connection than even principle theories with the concrete constituents of matter like atoms or particles: information-theoretic device-independent theory does not presuppose any kind of physical system at all.

Should one take quantum theory to be a theory of (a particular kind of) information? Such proposals appeared even before the advent of device-independent methods~\cite{bubstudies,mydiss}, while the latter give them a mathematical expression. Take the example of the Tsirelson bound~\cite{Tsirel}. If physics is captured by the probabilities $\mathbf{p}$, then all of quantum physics, including quantum bipartite correlations, must stem from some constraints on $\mathbf{p}$. Available constraints for the derivation of the Tsirelson bound are information-theoretic: a limit on communication complexity~\cite{van_dam_nonlocality_2000}, nonlocal computation~\cite{linden_quantum_2007}, the possibility of a well-defined classical transition (macroscopic locality)~\cite{Masanes}, or information causality~\cite{pawlowski_information_2009}. Whichever assumption one chooses, a non-trivial result is that quantum mechanics emerges in a purely information-theoretic context. It is legitimate to wonder whether such a theory is still affixed to reality and if yes, in what sense.

\section{Relation to realism}\label{sect_onto}

In a well-known argument purporting to show incompleteness of quantum mechanics, Einstein proclaimed that quantum theory would be complete if the wavefunction $\psi$ described ``the real state of the real system''~\cite{Einstein_let1}. While Bell inequalities were used to attack the reality of states, device-indepen\-dent methods do away with the idea of real systems. It is conceivable that in the early 1930s Einstein did not even contemplate such a possibility. Consider the opening lines of the EPR article written by Podolsky~\cite{HowardEinstein}:
\begin{quote}
Any serious consideration of a physical theory must take into account the distinction between the objective reality, which is independent of any theory, and the physical concepts with which the theory operates. These
concepts are intended to correspond with the objective reality, and by means of these concepts we picture this reality to ourselves.~\cite{EPR}
\end{quote}
A similar dictum can be found in Dirac's 1930 textbook of quantum mechanics, with `objective reality' replaced by the philosophically unambitious `physical entities':
\begin{quote}
``The most powerful advance would be to perfect and generalize the mathematical formalism that forms the existing basis of theoretical physics, and after each success in this direction, to try to interpret the new mathematical features in terms of physical entities.''~\cite{Dirac}
\end{quote}
Device-independent approaches render both philosophies obsolete. If a theory contains no notion of system, there is no reason to picture reality as comprised of physical entities. For sure, device-independent physics still informs us about Nature but, in a uniquely radical rejection, it does not support a claim that either Nature or physical theory are constituted of entities. This is more powerful than the widespread withering away of entity realism~\cite{French98}, a statement that physical entities are objectively existing, real things. Systems in the device-independent approach are unnecessary not only for the purposes of interpretation, but also on the theoretical side. They cannot correspond to objective reality because they are absent from the theory. Both in the philosophy of physics and in its mathematics systems are no more a requirement.

Device-independent methods promote a view that is also more powerful and unusual that the rejection of `naive' realism, which continues to characterize many working physicists. Naive realism is an uninformed form of entity realism stating that the objects of experimental science, think electrons or photons, are real because the empirical work and the laboratory heuristic suggest so. First the wave-particle duality, then Heisenberg's indeterminacy relations and the Kochen-Specker contextuality removed all possibility of a consistent account along these lines. Device-independence runs contrary to the experimental heuristic of naive realists to such extent that achieving it in the laboratory becomes a serious challenge. Boxes are usually built out of known systems like photons, yet no knowledge of such systems can be supposed by the experimenter, or the setup would immediately turn into a device-dependent one. That experimentalists often leave unnoticed minor device-dependent assumptions shows how counter-intuitive device-independent physics can be for a naive realist.

A view of reality that comes closest to accommodating device-in\-depen\-dent approaches was expounded by John Wheeler in an unpublished notebook~\cite{wheel74}:
\begin{quote}
On this view physics is not machinery. Logic is not oil occasionally applied to that machinery. Instead everything, physics included, derives from two parents, and is nothing but cathode-tube image of
the interplay between them. One is the ``participant.'' The other is the complex of undecidable propositions of mathematical logic.~\ldots~The propositions are not propositions about anything. They are the abstract building blocks, or ``pregeometry,'' out of which ``reality'' is conceived as being built.
\end{quote}
Wheeler's ``machinery'' seems to correspond to Einstein's notion of constructive theory, and Wheeler in the notebook contends against its prevalent position among practising physicists. The entire excerpt could be read as an argument in favour of principle theories had it not been for one key sentence: propositions, according to Wheeler, do not refer to ``anything'' and have no semantics, yet they are the building blocks of a conception of reality. This conception, while being ``as full-blown as anyone could want''~\cite{Fuchs2016}, stems from an abstract model in a direct consequence of Wheeler's ``participatory'' philosophy. Even someone whose philosophical standpoint may be entirely different cannot help noticing a curious occurrence, perhaps a first in the philosophy of physics, of a view that the propositions are themselves elements of reality and that they do not need to refer to any entities whatsoever, whether empirical or theoretical. Device-independent models proceed on a similar view replacing Wheeler's ``undecidable propositions'' by an ensemble of operationally defined inputs and outputs.

\section{Example: ``Almost quantum'' cor\-rela\-tions}\label{sect_almost}

Some constraints on $\mathbf{p}$ allow for an interpretation of the device-in\-depen\-dent setup in terms of systems. This \textit{emergence} of systems needs to be demonstrated mathematically, and an important check is the type of composition rule for such emerging entities. Quantum theory describes composition via the tensor product structure. If one posits that the no-signalling box is described by a global Hilbert space, one needs to test the availability of the tensor product between the subspaces that characterize each party. The work on the so-called ``almost quantum correlations'' addresses this question~\cite{navascues2007,navascues2015almost}.

Remarkably, there is enough leeway between two assumptions: the existence of the global Hilbert space and the tensor product structure of local subspaces. Rather than by the tensor product, the condition of independence of local observables can be captured by commutativity between two families of projectors pertaining to different parties. Correlations exhibited by the models based on commutativity relations differ slightly from quantum correlations: they are ``almost'' quantum.

The notion of subsystem, and with it the notion of physical system, is put into question in the device-independent models leading to almost quantum correlations. This is apparent in the definitions given by several authors working on this topic. It is not unusual to find a common-sense expression of device-independence in the familiar language of local subsystems:
\begin{quote}
Consider a scenario where $n$ parties conduct measurements $\bar{x}=(x_1,\ldots,x_n)$ on \textit{their respective subsystems}, obtaining outcomes $\bar{a}=(a_1,\ldots,a_n)$.~\cite[our emphasis]{navascues2015almost}
\end{quote}
The notion of subsystem involved is, however, different from the usual one. Rigorously speaking, it has to be defined algebraically:
\begin{quote}
Subsystems are defined by specifying observable algebras: these are assumed to be $C^*$-algebras that mutually commute.~\cite{Fritz}
\end{quote}
According to common sense, this algebraic definition must be a mere rephrasing of the usual Hilbert space notion. To check this, one appeals to the composition rule. It transpires that the result is negative: the notion of subsystem in the sense of commutativity of subalgebras does not correspond to the usual idea of physical systems that are statistically independent~\cite{florig}. Fritz formulates a conceptual lesson:
\begin{quote}
  It is our point of view that the operation of forming a composite system $\mathcal{H}_A\otimes \mathcal{H}_B$ from its subsystems $\mathcal{H}_A$ and $\mathcal{H}_B$ should not be a fundamental structure in a physical theory. The point is that nature presents us with a huge quantum system which we observe and conduct experiments on, and in some ways this total system behaves as if it were composed of smaller parts. Hence it seems that the correct question would be ``When does a physical system behave like it were composed of smaller parts?'' rather than ``How do physical systems compose to composite systems?''. Note that this is in stark contrast to many other approaches to the foundations of quantum theory, in which the operation of forming a composite system from subsystems is a fundamental structure.~\cite{Fritz}
\end{quote}
Thus systems in the framework of almost quantum correlations do not obey ordinary intuition. Even if, as in this example, the new mode of explaining physical phenomena occasionally refers to systems, one must remain aware of the pitfalls. A common-sense notion of system is clearly unhelpful, yet it is no accident that the authors strongly desire that their framework be explanatory of physics: ``The ubiquity of the almost quantum set $\tilde{Q}$ [\ldots] seems to suggest that it emerges from a reasonable (yet unknown) \textit{physical} theory''~\cite[our emphasis]{navascues2015almost}. If such future theory is not a theory of physical systems, what would it be a theory of? One finds a tentative answer in a definition using only the strictly necessary concepts:
\begin{quote}
For Alice (respectively for Bob), an experiment is a process or black box to which she feeds an input $x$ from the alphabet $\mathcal{X}$ and from which she receives an output $a$ from the alphabet $\mathcal{A}$. Alphabets $\mathcal{X}, \mathcal{Y}, \mathcal{A}, \mathcal{B}$ are of finite cardinality.~\cite{Lang}
\end{quote}
On this view, physical theory is about languages: it is defined by a choice of alphabets for the inputs and the outputs and by the conditions imposed on this algebraic structure. Strings, or words in such alphabets, form a common mathematical background of device-independent approaches.

\section{Physical theory is about languages}\label{sect_lang}

The origin of the idea that in the foundation of quantum theory should lie an abstract mathematical model can be traced back to von Neumann's first publication on quantum mechanics~\cite{HvNN}. Prior to the advent of device-in\-depen\-dent approaches, its meaning was perhaps best expressed by Wheeler. According to his formulation, a mathematical model should contain as few assumptions as possible, including the removal of presuppositions about the Hilbert space, which may reemerge in the course of a posterior ``treatment'':
\begin{quote}
2. Start with what formal system?
\\
Take a formal system. Enlarge it to a new formal system, and that again to a new formal system, and so on, by resolving undecidable propositions (``act of participation''). Will the system become so complex that it can and must be treated by statistical means? Will such a treatment make it irrelevant, or largely irrelevant, with what particular formal system one started?~\cite{wheel74}
\end{quote}

The conclusion drawn at the end of the previous section, \textit{viz}., that physical theory is about languages, belongs to Wheeler's methodological lineage, although one will likely learn more about the relation between individual ``treatment'' and mathematical formalism as other device-in\-depend\-ent models come to light. The meaning of `physical'---a term Wheeeler did not use in his notes---usually implies a semantic reading in the form of a statement that the theory `describes' some entities or is `about something.' But the examples given in Sections \ref{sect_causal} and \ref{sect_almost} support a new reading of Wheeler's statement that ``propositions are not about anything.'' On this view, device-independent physical models are not `about' systems. A semantic interpretation, or `aboutness,' although not necessarily wrong, is adduced at a later stage and sometimes proves unhelpful for understanding physics. Remove it: now a new interpretation need to be given to the bare mathematical formalism.

If, as Wheeler said, physical theory ``is not about anything''---not about \textit{any thing}---what is it about? It is tempting to answer, as hinted in Section~\ref{sect_reconstruct}, that physical theory is about information or a special kind thereof. Not all conceptual problems, however, get so resolved. If information is a more fundamental substance, does it come in many kinds or varieties? One possibility would be that such types of information are all similar in structure (obeying \textit{the} concept of information, e.g., as defined by Shannon~\cite{shannon}) but vary in the values of some parameters. Another option is to radically distinguish one notion of information (e.g., information that cannot be cloned) from all others~\cite{bub_why_2012}.
But if quantum theory is about information that cannot be cloned, one is left wondering who is the agent trying to clone it. This interpretation still leaves open the problem of observer who `stores' and `handles' information.
In our view, such conundrums are misleading, because the term `information' is not required to drive home the point of device-independent approaches. `What is physical theory about?'---it is only appropriate to search for an answer by looking at the mathematical formalism of device-independent methods. What needs to be understood, therefore, is the common conceptual background of the various mathematical constraints on $\mathbf{p}$. Such a background should become a common philosophical denominator of physics in lieu of Shannon's or von Neumann's information theory. It should also leave no open interpretational gaps such as a reference to a meta-theoretic agent.

Section~\ref{sect_almost} gives an example of a physical model based on strings in finite alphabets. In a device-independent approach, only an empirical interpretation of such strings is readily available: they are the operationally defined input-output records. As in Shannon's theory of information, it is possible to study them in a purely formal way irrespectively of their semantics. The entities that the strings are purportedly `about' are irrelevant constructs and, even when they are referred to under the name of `systems,' they have no bearing on how the construction of the theory proceeds mathematically.

Wheeler's methodology provides guidance: at the beginning, ``take a formal system.'' In a rebuttal of his position, one could argue that a semantic view of device-independent models is necessary if the theory is to be physical rather than purely mathematical: physical, hence somehow relating to nature. A minimal requirement would be to link with empirical data. Wheeler's ``not about anything'' can be circumvented by keeping the preposition ``about'' while clearing oneself of ``any thing'': in the absence of semantics, the only available interpretation is `about languages.' If strings are not `about' some elements of reality, they can be said to be `about' languages from which they are formed. These languages are the only remaining theoretical constituents, making the `about-languages' reading a minimal interpretation one level below `about-systems.' The emphasis on languages rather than physical entities is not new in the philosophy of physics. Niels Bohr underlined the role of language:
\begin{quote}
Bohr went on to say that the terms of discussion of the experimental conditions and of the experimental results
are \textit{necessarily} those of `everyday language,' suitably `refined' where necessary, so as to take the form of classical dynamics.~\cite[p.~38]{bohm}.
\end{quote}
While Bohr's attention was focused on everyday language and classical dynamics, the connection he identified between them is instructive: a `refinement' of a linguistic structure produces a physical description. The same methodology applies to device-independent models, bar the replacement of everyday language by a formal notion of language. A set of conditions then `refines' this concept producing a full-blown physical theory.

A sketch of a purely formal study of strings that form the foundation of physical theory runs as follows, bearing a slight resemblance to Wheeler's suggestion that a particular formal system one started with may become ``largely irrelevant.'' The constrains imposed on the set of all possible strings pick out a subset of allowed words. A valuation is introduced on this set, which enables the use of the probability calculus. Its meaning can be read on many levels: (a) as a prediction of measurement results; (b) as an update of the observer's information; (c) as emergent causal relations between events; (d) as a correlation between particular inputs and outputs. As in Shannon's theory, the choice of interpretation has no bearing on mathematical results. Their minimal conceptual import is that they mean no more than what they are: an algebraic, or a linguistic, structure. Geometric constructs, including the Hilbert space of quantum theory, do not underwrite this minimal reading as they tend to be device-dependent. In a device-independent approach, the notion of state space is absent; this, in turn, removes the necessity of pointing to the system which this state space describes. Physical theory, on this minimal reading, is about an algebraic structure formed by the strings obeying certain conditions: a view of a physical theory about languages.

This view has precedents in the history of quantum theory~\cite{grin2015}. In the 1950s, Hugh Everett argued that the observer is characterized by memory, i.e., ``parts... whose states are in correspondence with past experience''~\cite{everett}. A modern, functional definition of memory is that it serves to store strings. Everett's understanding of the observer is, then, in line with a view that relies on strings as fundamental elements of physical theory.
Several decades later, Wojciech Zurek suggested that algorithmic complexity of the observer's description of the state of the system ought to be added to physical entropy~\cite{Zurek1,Zurek3}. His interpretation involved full-blown quantum theory. This need not be so: the observer's information can, in a reversal of Zurek's strategy, be viewed as a fundamental ingredient without any semantic content. Mathematically, once again, it only involves strings in a finite alphabet in which the observer operates.

Device-independent approaches exacerbate the need for a minimal conceptual reading to support the mathematical formalism. An algebraic foundation leads to a common philosophical denominator of the operational models focused on the inputs and the outputs, which only relies on the abstract content of the theory. An extended interpretation would adduce `systems' as remnants of the old explanatory mode; a minimal one removes all semantic content but it is also more amenable to mathematical analysis that the vaguer ``physics is about information.'' While, mathematically speaking, information comes in only two varieties: Shannon and von Neumann, the statement that ``physical theory is based on languages'' opens up a multitude of choices among the formal structures of algebraic coding theory. It introduces new tools yet unused in physics; what is more, unlike the informational interpretation, it does not refer to a meta-theoretical agent. Similarly to the creation of an abstract, non-semantic theory of information by Shannon, a device-independent model based on languages provides a minimal but sufficient ground for physical theory.

\section{Conclusion}
In quantum cryptography, it has always been allowed, even customary, to ask the anathema question of physical theory: what if a preparation or a measurement device is cheating on the experimenter? Is it still possible to obtain meaningful results? Under the influence of cryptography, quantum theory developed a way of doing physics that can accommodate such questions: a device-independent approach. Conceptually, device-independence does not require that the notion of system be present in physical theory. It is then legitimate to ask what such a new physical theory is about. Device-independent models of indefinite causal orders and almost quantum correlations suggest a possible answer: it is about languages. Like information-theoretic postulates that lead to the derivation of quantum theory in the operational framework, particular constraints on languages produce a device-independent model that exhibits characteristic features of quantum theory. Further consequences of this novel view of physical theory remain to be explored both conceptually and mathematically.

\section*{Acknowledgements}

Many thanks to \v Caslav Brukner, Chris Fuchs and Tom Ryckman for helpful comments. Part of this work was supported by FQXi under grant FQXi-MGA-1505.

\singlespacing\footnotesize

\bibliographystyle{habbrv}

\end{document}